\documentclass{iau}
\usepackage{amsmath}
\usepackage{graphicx}
\usepackage{multirow}

\begin{document}

\lefttitle{Tomek Kamiński}
\righttitle{Bipolar red nova remnants}

\journaltitle{Planetary Nebulae: a Universal Toolbox in the Era of Precision Astrophysics}
\jnlDoiYr{2023}
\doival{10.1017/xxxxx}
\volno{384}

\aopheadtitle{Proceedings IAU Symposium}
\editors{O. De Marco, A. Zijlstra, R. Szczerba, eds.}
 
\title{Red novae, stellar mergers in binary and triple systems, and bipolar nebulae
}

\author{Tomek Kamiński}
\affiliation{Nicolaus Copernicus Astronomical Center, Polish Academy of Sciences, Rabiańska 8, 87-100 Toruń, Poland}

\begin{abstract}
Red novae are transients powered by collisions of non-compact stars. Among their progenitors are systems of evolved subgiants and giants stars. Remnants of such red novae display bipolar structures which have remarkably close characteristics to many post-AGB or pre-PN systems. It is important to ask (and eventually verify) whether some of the less known post-main-sequence objects (mis-)classified as pre-PNe can be merger remnants similar to the red nova remnants.  
\end{abstract}

\begin{keywords}
stars: imaging, stars: binaries, stars: winds, outflows, circumstellar matter
\end{keywords}

\maketitle

\section{Red novae as stellar mergers}
Red novae, also known as luminous red novae, represent a subset of eruptive stars resulting from stellar mergers, particularly non-compact stars. This group includes five Galactic objects: CK Vul, V4332 Sgr, V838 Mon, V1309 Sco, and EWS 2002-OGLE-BLG-360. Unlike supernovae and classical novae, red novae primarily derive their energy from accretion \citep{TS2006}, not nuclear processes, leading to lower temperatures of their immediate remnants. The stellar products of red novae are cool giants or supergiants with temperatures of 2000--3000 K, surrounded by complex circumstellar environments rich in cool gas and dust \citep{KamiALMACK1,Kami2018}.

Despite being a relatively small sample, Galactic red novae offer insights into a diverse range of variables. The timescales for studying these remnants typically span decades, with CK Vul's outburst observed as early as 1670. Extragalactic red nova analogs in the Local Group, while numerous, lack sufficient spectral observations; observations of the merger products and of their progenitors are also sparse \citep{Pastorello}. 

Red novae are believed to be mergers involving non-compact stars, including solar-mass giants/subgiants \citep[e.g.,][]{Stepien} and main-sequence stars \citep[e.g.,][]{TylProgV838}, but the collision in CK Vul involved a red-giant-branch (RGB) star with a He white dwarf  \citep[WD;][]{TylKami2024}. These events might also encompass planetary engulfment \citep{De,SokerPlanets,Engulf}. The luminous extragalactic objects tend to involve higher mass stars (say 3--10\,M$_{\odot}$ or higher), often in the Hertzsprung gap \citep[e.g.][]{Nadia101,Nadia-bwo}. The diversity of binary configurations is large in the context of stellar evolution and stellar masses involved, but collectively, red novae have introduced a unique new opportunity to investigate the intricate physical mechanisms destabilizing binaries and resulting in extreme stellar interactions. There is hope that these transients and their remnants will aid our understanding of the common envelope phase \citep{Ivanova}.

Distinguishing characteristics of red novae include their eruptive durations lasting from a few months to a year, intermediate luminosities ($\lesssim10^6$ L$_{\odot}$) between classical novae and supernovae, multipeaked light curves, and rapid cooling to low photospheric temperatures (1800--2000 K) after the eruption. This abrupt cooling sets red novae apart from classical novae, which progress to the coronal phase at $10^5$ K. The cooling process facilitates efficient formation of molecules and dust, processes which are relatively easy to observe but not easy to model \citep[e.g.,][]{Iaconi,morgan22}. There are other Galactic objects thought to be of a similar nature as red novae (mergers), including for example objects like the Blue Ring Nebula \citep{BlueRing} or Phoenix giants \citep{Melis}, but because there is no record of their outburst (``mergeburst''), they are not part of the red nova classification. It is hard to obtain  realistic statistics of these transients but the best efforts predict two bright red novae per decade per galaxy \citep{Kochanek,Howitt,stats2023}. Red novae are considered to be easily discoverable by the Vera C. Rubin Legacy Survey of Space and Time \citep[LSST;][]{LSST}. As no new Galactic red nova has been observed since 2008, an overdue event is in the offing. A search for red nova progenitors is ongoing as well \citep{progSearch}.  

A fuller review of Galactic red nova properties was given in \citet{Kami2018}.

\section{Bipolar structures in red nova remnants}
\subsection{Coalesced binaries}
\begin{figure}
\begin{center}
 \includegraphics[width=7cm]{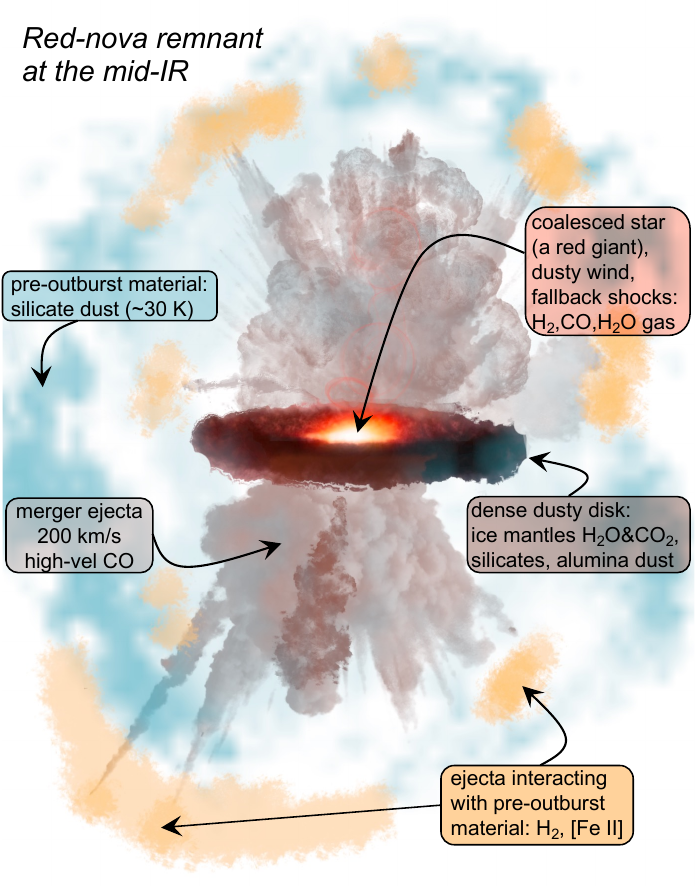} 
\end{center}
\caption{A schematic representation of the complex architecture of a red nova remnant. Observable features are labeled. They are not in scale. The scheme matches well observation of V1309 Sco and V4332 Sgr discussed in the text.}
\label{fig1} 
\end{figure}

Recent studies of the Galactic objects emphasize the complexity of circumstellar envelopes, suggesting the formation of disks, jets, wide-angle outflows, and inhomogeneous winds in the aftermath of mergers \citep[see][for the most recent discussions]{Mobeen,Steinmetz}. This complex architecture is illustrated in Fig.~\ref{fig1}. The bipolar structures discovered in red nova remnants are most relevant to the connection to planetary and pre-planetary nebulae. 

One important example of such a bipolar circumstellar remnant is V1309 Sco which erupted as a red nova in 2008. V1309 Sco stands out as a key case supporting the merger scenario for red novae due to its well-documented variability before the 2008 eruption. Photometric data revealed V1309 Sco as an eclipsing contact binary with an exponentially decreasing orbital period, showcasing the spiraling-in of two non-compact stars \citep{Tyl1309}. The subsequent release of gravitational energy and shocks directly triggered the red nova flash \citep{MP2017}. Optical, infrared, and  submillimeter observations a few years to decade after the outburst display characteristics which can only be explained by a highly bipolar structure \citep{Steinmetz}. Spatially unresolved spectral observations of the V1309 Sco's remnant display two fast outflows which produce shocks of different excitation and chemical properties --- one giving rise to H$_2$ IR emission, the other hinted by emission of HCO$^+$ through pure rotational transitions. Additionally, emission of molecular gas at lower gas temperatures was mapped directly in rotational lines of CO and SiO by the Atacama Large Millimeter Array (ALMA) interferometer \citep{Kami2018,Steinmetz}. These maps show that the merger ejecta has a bipolar structure, but whether it is caused by wide-angle outflows or a pair of well collimated jets is currently unclear, as the ALMA observations only barely resolve the source. Observations with ALMA at higher angular resolutions will certainly be feasible in the future and should be more reveling, especially since the source is rapidly expanding.       

V4332 Sgr is the older twin of V1309 Sco. It erupted in 1994 and so far has shown spectral and photometric behavior remarkably similar to V1309 Sco \citep{KamiV1309}. It is thus believed that V4332 Sgr originated from a merger of evolved (post-MS) solar-mass stars, but, unlike in V1309 Sco, no direct observations of the progenitor system are available. Nevertheless, the remnant of V4332 Sgr has been much better resolved by ALMA and a three-dimensional model of its cool component was constructed \citep{Kami2018} with the Shape-X software \citep{shape}. The study reconstructs the remnant as a fully bipolar structure of $\sim$600 AU expanding according to the Hubble law (i.e., expansion velocity is proportional to the distance from the expansion center). The model architecture is very similar to bipolar planetary nebulae and, even more, to young pre-planetary nebulae, but the amount of detail in V4332 Sgr is very limited by the angular resolution which loses the fight against the large distance to the source of 4--6 kpc. \citet{Kami2018} interpret the structure of V4332 Sgr's  circumstellar medium as a result of outflow which was deflected into bipolar lobes by a pre-existing dense disk or torus formed in the orbital plane of the evolving binary. This scenario gains a lot of support in merger simulations of \citet{Pejcha2017}, which reproduce well the general shape of the circumstellar remnant and its kinematical structure. The observations of V1309 Sco and V4332 Sgr seem to suggest that the bipolar structure arises as an intrinsic feature of the merger event and is directly linked to the parameters (e.g., orientation) of the decaying binary.   

\subsection{Mergers in triple systems?}
Mergers are likely to take place in triple and higher systems \citep[e.g.,][]{Glanz,Hamers}, for instance, as a consequence of Kozai-Lidov cycles. The most famous example of a red-nova progenitor which was triple is V838 Mon \citep{TylProgV838,KamiALMAV838}, and {\it source I} in the Orion Molecular Cloud 1 may be an example of a merger in a multiple system \citep{sourceI}. Both objects are very young, while we are interested here in evolved objects which may shed light on our understanding of planetary and pre-planetary nebulae. It happens that the oldest known red nova, CK Vul, was likely a triple system. 

CK Vul, initially thought to be the earliest documented nova \citep{Shara}, is a unique object with a complex history. Discovered in 1670, CK Vul has defied classification as a typical nova due to its peculiar light curve, displaying multiple peaks, and a reddish color, both features --- as understood now -- characteristic of red novae \citep{KamiNature}. The object's nature became clearer with modern observations, lending a large body of evidence for coalescence of a He white dwarf (WD) and a red giant branch (RGB) star \citep{Tyl2024}. The evidence is most compelling in the elemental and isotopic composition of the circumstellar gas \citep{KamiSingleDish}, especially due to the presence of AlF molecules containing the radioactive isotope of $^{26}$Al \citep{KamiAl}.   

\begin{figure}
\begin{center}
 \includegraphics[width=12cm]{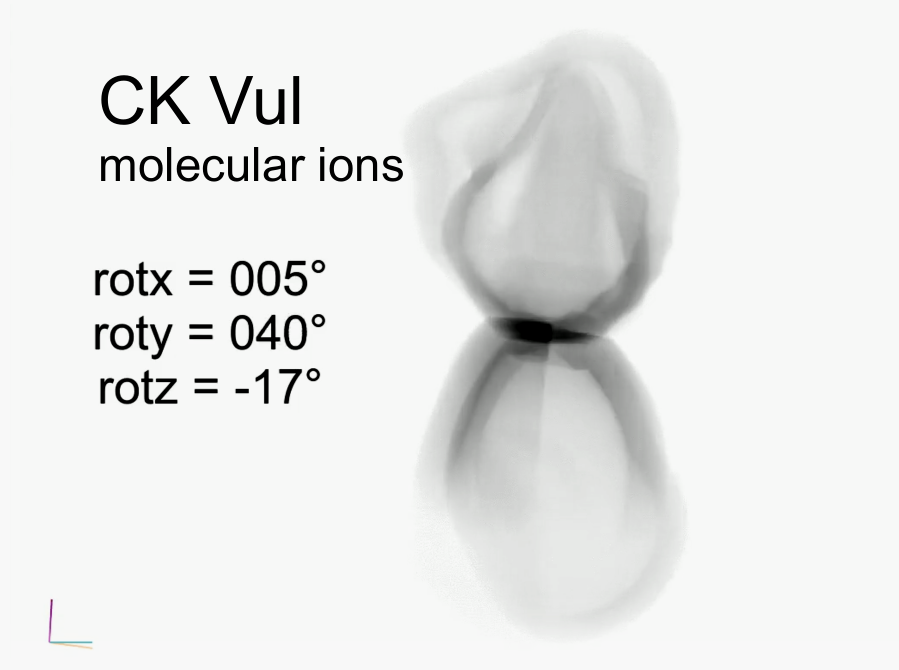} 
\end{center}
\caption{Reconstructed distribution of molecular ions in the remnant of CK Vul. Adopted from \citet{KamiCKII}. The remnant has a very complex structure when viewed in three dimensions.}
\label{fig2} 
\end{figure}

This hypothesized stellar collision led to the ejection of material, forming a large hourglass nebula with multiple bullets and a dust-rich waist \citep{hajduk1,hajduk2,Banerjee}. The large nebula is mostly traced in emission of recombining plasma and can easily be taken for a planetary or pre-planetary nebula (if not the bizarre central object). Studied through millimeter and submillimeter observations, in particular with the Submillimeter Array and ALMA, CK Vul also revealed a smaller and cooler molecular component \citep{KamiNature,KamiCKI,KamiCKII}. While the large hourglass and the bullets very likely originate in the 1670--1672 events, the age of the molecular component remains practically unconstrained. With detailed spatio-kinematic maps (i.e., spectral cubes) of emission from neutral and ionic molecular species, \citet{KamiCKII} constructed thorough three-dimensional models of the remnant. The observations and models of the CK Vul remnant are of a much better quality than those for V1309 Sco and V4332 Sgr discussed above, in part because CK Vul is much older and the expansion velocities are twice higher there. A sample view of the reconstructed structure of CK Vul is presented in Fig.~\ref{fig2} and a full movie can be seen at \url{https://www.aanda.org/articles/aa/olm/2021/02/aa39634-20/aa39634-20.html}. Based on these models, the molecular gas exhibits an overall bipolar structure and intricate three-dimensional architecture which exhibits strong point symmetry. Shocks play an important role in excitation of the molecular gas and determine its chemical composition (including the formation of complex organic molecules like methanol or methylamine). The analysis strongly suggests multiple ejections with varying inclinations and position angles. It is unlikely that these were created during the 1670--1672 outburst. One possible interpretation is that some material originally ejected during the mergeburst falls back on the coalesced star causing episodic accretion and (feedback) mass loss through collimated streams. However, it is unclear whether such a mechanism could ensemble a cool molecular nebula of significant mass (see below). Another interpretation is that CK Vul was a triple system before the merger, and the surviving companion interacts with the merger ejecta. This is a highly speculative scenario but explains more naturally an observed misalignment of the main axis of the large hourglass-shaped recombining nebula with respect to the axis of the molecular lobes.   

With submillimeter observations of CO and under some assumptions on CO abundance, one can calculate the mass of the lobes. \citet{KamiCKII} found a mass of at least 0.8\,M$_{\odot}$. This is a very high mass when compared to simulations of stellar mergers where only a few percent of the mass of the entire stellar system is dispersed. This mass is, however, not much larger than masses of envelopes of some pre-planetary nebulae. Moreover, the linear momentum of $10^{40}$ g cm s$^{-1}$ and the kinetic energy of $10^{47}$ erg stored in the molecular outflows of CK Vul are comparable to those of many pre-planetary nebulae \citep[cf.][]{Valentin}.  
\section{Curious relation of red nova remnants to pre-PNe}
I have highlighted the many correspondences between the merger remnants and (some) pre-planetary nebulae (PPNe). There is a possibility that mechanisms responsible for creating the red nova remnants are similar, if not the same, as in PPNe. Binary interactions and common-envelope ejections certainly play a crucial role in the formation of both classes of objects, and it is worth studying how far the parallels go. Some authors go as far as to suggest that merger and related events could have created some of the well known bipolar planetary nebulae \citep{SK}. Since some bipolar flows in PPNe must be created in a short time (often shorter than a timespan of years) and with an extra source of energy other than the radiative energy of the central star, a merger event is a scenario worth considering for each of these cases. The remarkable commonalities in the {\it appearance} of the circumstellar media of these objects might also mean that some of the sources classified today as evolved stars with the PPN status can in fact be merger products. Whether some particular PPNe could be interpreted as stellar collision products have been discussed in the past, e.g., for Frosty Leo \citep[;][]{Bourke}, the Rotten Egg Nebula, or PN K4-47 \citep{Debra}. One important difference between the discussed red novae remnants discussed here and PPNe is a low luminosity of the central object, which is the case of CK Vul is 12--60~L$_{\odot}$. These are much lower than 10$^4$~L$_{\odot}$ expected for any post-AGB star. This distinguishing feature is, however, of limited use for objects of unknown distance and total extinction, which unfortunately is a frequent case.  Nevertheless, I remain hopeful that the next merger remnant will be recognized among PNe or PPNe.

\medskip
I acknowledge funding from the Polish National Science Center grant SONATA-BIS no. 2018/30/E/ST9/00398.

\end{document}